\documentclass[showpacs,aps,pra, longbibliography, 10pt,reprint,
]{revtex4-2}

\usepackage{bm}
\usepackage[breaklinks=true,colorlinks=true,linkcolor=blue,urlcolor=blue,citecolor=blue]{hyperref}
\usepackage{dcolumn}
\usepackage{amsmath,amssymb}
\usepackage{mathdots}
\usepackage{natbib}
\usepackage{soul,color}
\usepackage{graphicx}
\usepackage[note-name]{notes2bib}
\usepackage{mathptmx}
\usepackage[normalem]{ulem}

\usepackage[T1]{fontenc}

\begin{document}

%\title{Vortex-Based Grating Coupler for Short-Wavelength Spin Waves}

\title{Higher-harmonic spin-wave excitation by a superconducting antenna with vortices}
%Vortex-Induced Nonlinearity in a Superconducting Antenna\\ for Efficient Second-Harmonic Spin-Wave Excitation}% Force line

%Second-Harmonic Spin-Wave Excitation by a Superconducting Antenna with Vortices

%A Superconducting Vortex Antenna for Nonlinear Magnonics

\author{Anton Pokusinskyi$^{1,2}$}
\email{anton.pokusinskyi@tu-braunschweig.de}
\author{Paweł Gruszecki$^3$}
\author{Julia Kharlan$^{3,4,5}$}
\author{Oleksandr Dobrovolskiy$^{1,2,6}$}
\author{Jarosław W. Kłos${^3}$}

\affiliation{$^1$Cryogenic Quantum Electronics, Institute for Electrical Measurement Science and Fundamental Electrical Engineering, Technische Universit\"at Braunschweig, Hans-Sommer-Stra\ss e 66, D-38106 Braunschweig, Germany}
\affiliation{$^2$Laboratory for Emerging Nanometrology (LENA), Technische Universit\"at Braunschweig, Langer Kamp 6A-B, D-38106 Braunschweig, Germany}
\affiliation{$^3$Institute of Spintronics and Quantum Information, Faculty of Physics and Astronomy, Adam Mickiewicz University, Poznań, Uniwersytetu Poznańskiego 2, 61-614 Poznań, Poland
}
\affiliation{$^4$V.G. Barykhtar Institute NAS of Ukraine,  Kyiv 03142, Ukraine}
\affiliation{$^5$G. V. Kurdyumov Institute for Metal Physics of the NAS of Ukraine,  Kyiv 02000, Ukraine}
\affiliation{$^6$FLUXONICS---The European Foundry for Superconducting Electronics e.V., 38106 Braunschweig, Germany}

\begin{abstract}
The interaction between superconductivity and magnetism provides new routes for controlling nonlinear excitations in hybrid systems. While dc-driven Abrikosov vortices (fluxons) are known to interact with spin waves, the ability of confined, nearly harmonic vortex oscillations to generate higher-frequency spin-wave modes remains unexplored. Here, by combining time-dependent Ginzburg--Landau modeling with spectral analysis of the microwave near field, we demonstrate higher-harmonic spin-wave excitation driven by a superconducting microstrip antenna in the Shubnikov phase. Unlike conventional harmonic-generation mechanisms associated with nonlinear magnetization dynamics, the magnetic layer remains a linear receiver, whereas the superconducting antenna generates the nonlinear microwave field. We show that nearly sinusoidal vortex motion and the accompanying redistribution of superconducting screening currents produce higher harmonics that couple to spin-wave modes. This response arises from spatial-to-temporal transduction, whereby harmonic vortex motion and the associated redistribution of screening currents convert a spatially nonuniform magnetic-field profile into higher temporal harmonics. Magnetic flux quanta thus act as active nonlinear elements for frequency conversion, extending the accessible spin-wave frequency range in superconducting--magnonic hybrids.
\end{abstract}
\maketitle

\section{Introduction}
Magnonics exploits collective spin precessions---spin waves---and their quanta, magnons, as carriers for wave-based information processing, offering a route toward energy-efficient computing beyond charge-based electronics \cite{Chu22tom}. Recent advances in nanoscale magnonic circuits \cite{Wan24pra}, topological magnonics \cite{Cla22cmp}, and nonlinear spin-wave dynamics \cite{Fle23jap} have greatly expanded the capabilities of magnonic devices, creating a growing demand for efficient generation and detection of coherent spin waves at nanometer length scales \cite{Pir21nrm}. Microstrip and coplanar waveguide microwave antennas are standard tools for the inductive excitation of spin waves \cite{connelly21}. However, efficient excitation of short-wavelength, high-frequency magnons requires antenna widths comparable to the spin-wave wavelength. Scaling metallic antennas to these dimensions leads to impedance mismatch with standard $50\,\Omega$ microwave circuits, increased ohmic losses, reduced excitation efficiency, and, because of the high current densities involved, substantial Joule heating that limits the maximum applicable microwave power \cite{Erdelyi25}.

Superconductors (SCs) provide an alternative to conventional metallic antennas owing to their exceptionally low microwave losses \cite{Mansour02,Liu19}. While SCs require cryogenic temperatures, this regime is increasingly relevant for magnonics, with spin-wave transport and dynamics attracting growing interest from liquid-helium to millikelvin temperatures \cite{Loo18pra,Zar22phr,Sch25prb}. In the Meissner state, where magnetic flux is expelled from the SC bulk, dc transport currents are dissipationless, while ac losses at GHz frequencies are orders of magnitude lower than in normal metals \cite{Dre13cmp}. In the Shubnikov (mixed) state, relevant to most technologically important type-II SCs, magnetic flux penetrates the material in the form of quantized flux lines -- Abrikosov vortices, or fluxons \cite{Bra95rpp}. The dynamics of vortices fundamentally modify the microwave response of an SC and introduce additional dissipation. Under dc transport currents, losses arise from vortex depinning and flux-flow motion, whereas under ac excitation they originate from the oscillatory motion of vortices within or beyond their pinning potentials \cite{Pom08prb,Dob20pra}, together with the finite surface impedance associated with quasiparticles (unpaired electrons) \cite{Vis11prl}. Yet, the microwave surface resistance of SCs remains orders of magnitude lower than that of normal conductors \cite{Tinkham1996}. Consequently, SC antennas can sustain high current densities with greatly reduced power dissipation, enabling the efficient generation of strong microwave magnetic fields at submicrometer scales.

Beyond their low microwave losses, SCs possess electrodynamic degrees of freedom absent in normal metals. Whereas conventional metallic antennas exhibit an essentially linear response in the regime considered here, the response of an SC additionally reflects the dynamics of the SC condensate \cite{Tinkham1996,Dre13cmp}, giving rise to intrinsic nonlinearities \cite{Ji89,Gurevich00}. In the mixed state, the response is further enriched by ac-driven Abrikosov vortices \cite{Beek93,Eley21} whose oscillations produce a time-dependent stray magnetic field with properties inaccessible to conventional antennas. In addition, vortex motion redistributes the transport and Meissner screening currents, modulating the antenna's microwave magnetic field \cite{Memarzadeh26}.

Recently, the interplay between superconducting screening, vortices, and spin-wave dynamics in SC/FM hybrids, where FM denotes a ferro- or ferrimagnetic layer, has attracted considerable attention  \cite{Bar96prl,Dob19nph,Gol19adm,Nie23pra,Kim18prl,Dob22mmm,Gol18afm,Jeo19pra,Bor23sci,Kha24sr,Zho24prb,Yut26phr,Ghi26mmm}. Prominent effects include SC-induced modifications of ferromagnetic resonance and magnetization dynamics \cite{Gol18afm,Jeo19pra}, vortex-mediated magnon scattering and transport \cite{Nie23pra,Kim18prl,Dob22mmm}, and the formation of hybrid spin-wave--Meissner-current modes \cite{Bor23sci,Zho24prb}. Moreover, rapidly moving dc-driven fluxons can excite spin waves via a Cherenkov-like mechanism when their velocity reaches or exceeds the magnon phase velocity \cite{Bes14prb,Phi23mmm,Dob25nan}.

By contrast, the coupling of spin waves to vortices undergoing small-amplitude, \emph{ac-driven} oscillations about a confined equilibrium position remains largely unexplored. Rather than traversing the SC strip, such vortices oscillate under the ac Lorentz force \cite{Coffey91} within a confining potential set by the finite strip geometry and the Bean--Livingston surface barriers at the lateral strip edges \cite{Bean64,Mik21prb}. Although the SC mixed state is intrinsically nonlinear, microwave-driven vortex trajectories can remain nearly harmonic over a broad range of excitation amplitudes \cite{Coffey91,Pom08prb}. This raises the question: Can nearly harmonic vortex oscillations generate higher harmonics in the local microwave magnetic field without an anharmonic time dependence \cite{Shk08prb} of the vortex displacement?

To address this question, we model the SC dynamics using the time-dependent Ginzburg--Landau (TDGL) equation and analyze the spectrum of the resulting microwave magnetic near field in an underlying magnonic conduit coupled through the stray fields. We show that an SC microstrip antenna in the mixed state excites spin waves at the second harmonic, with weaker contributions at higher harmonics. These components remain pronounced even when the vortex trajectories are nearly sinusoidal. We attribute this response to \emph{spatial-to-temporal transduction}: Vortex oscillations and the associated redistribution of SC screening currents convert the spatially nonuniform magnetic near field into temporal harmonics. In this mechanism, the FM layer thus acts as a linear receiver of the nonlinear microwave field generated by the SC antenna. The resulting local fields at $2f$, $3f$, and higher multiples of the drive frequency $f$ extend the frequency range available for spin-wave excitation without relying on nonlinear magnon dynamics \cite{Pir21nrm,Dre22nac}. In all, our results identify Abrikosov vortices as active nonlinear elements for frequency conversion in SC--magnonic hybrids, extending the accessible frequency range for spin-wave excitation in cryogenic magnonic devices.

\section{Geometry and materials considerations}

\begin{figure}[t!]
\centering
\includegraphics[width=8.6cm]{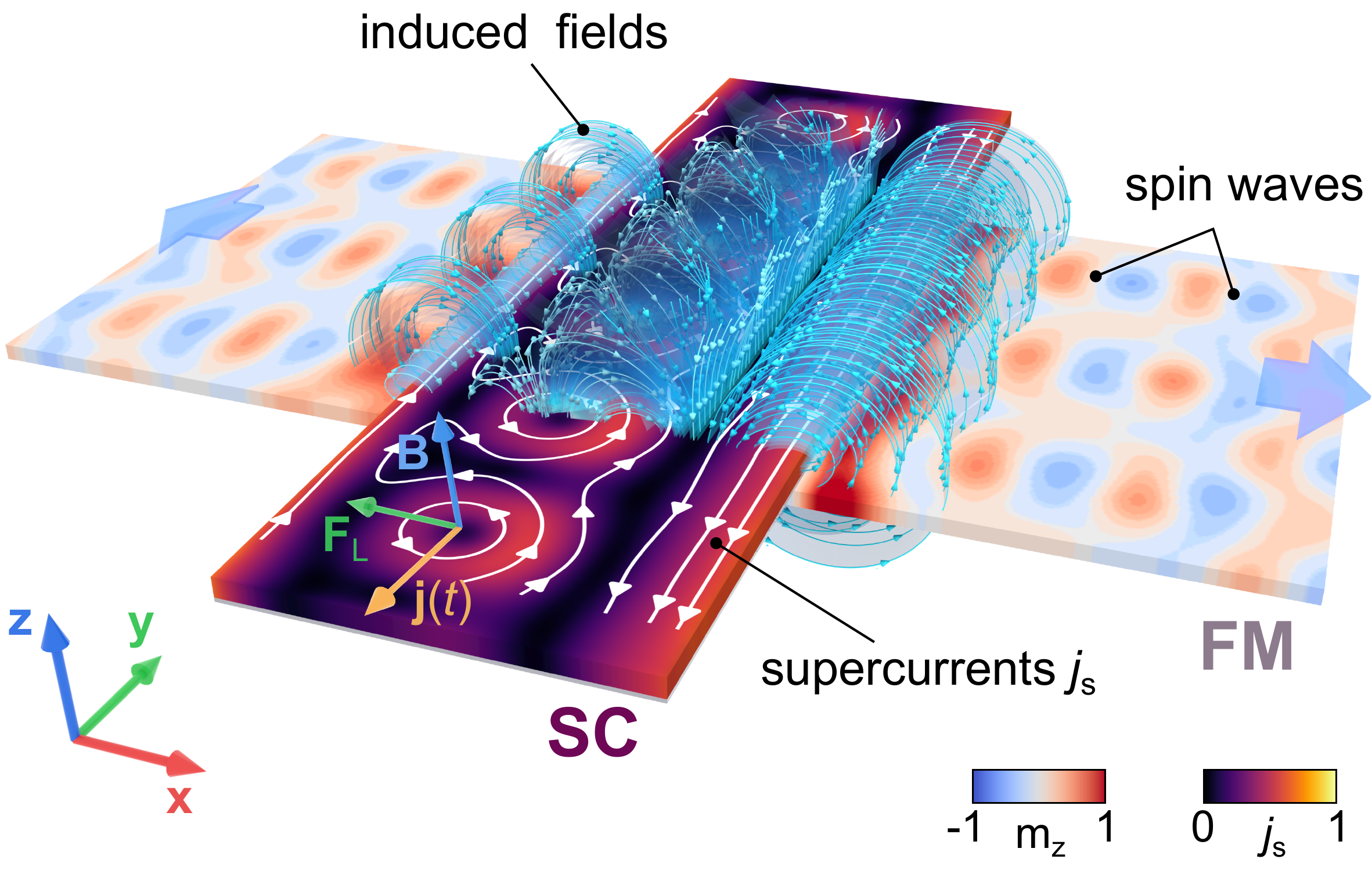}
\caption{
    Schematic of the considered geometry. A superconducting (SC) strip antenna in an out-of-plane magnetic field $B_0=30$\,mT carries a GHz-frequency ac current density $\mathbf{j}(t)$, whose Lorentz force $\mathbf{F}_\mathrm{L}$ drives transverse oscillations of Abrikosov vortices. Despite their nearly harmonic motion, the vortex stray field varies nonsinusoidally at a fixed observation point. This modulation, together with vortex-induced current redistribution in the SC, generates higher harmonics that excite spin waves in the underlying magnonic conduit (FM) with perpendicular magnetic anisotropy (PMA). 
    The dynamic out-of-plane magnetization $m_z$ is normalized to the saturation magnetization $M_\mathrm{s}$, while $j_\mathrm{s}$ represents the nonnegative magnitude of the superconducting current density, normalized to the amplitude of the applied ac current density for visualization.
}
\label{fig:system}
\end{figure}

We consider the system shown in Fig.~\ref{fig:system}: a SC strip is placed above a magnetic layer serving as a spin-wave conduit and separated from it by a 5\,nm-thick nonmagnetic insulating spacer. The two subsystems are coupled inductively through the microwave stray field \cite{Yut26phr}. An ac current at frequency $f$ is applied to the SC strip, which acts as a microwave antenna whose stray field excites spin waves in the magnetic layer. 

We consider two magnonic materials with perpendicular magnetic anisotropy (PMA): Ga-substituted yttrium iron garnet (Ga:YIG) and CoFeB. In both materials, PMA stabilizes an out-of-plane magnetized state supporting forward-volume spin-wave (FVSW) propagation. This configuration is particularly advantageous because the nearly circular magnetization precession suppresses conventional intrinsic nonlinear frequency-conversion processes within the magnetic layer. These processes are discussed in \textcolor{blue}{Appendix~A}. Ga:YIG serves as the primary magnonic platform in this study, combining relatively weak PMA with a substantially reduced saturation magnetization $M_\mathrm{s}$. As a complementary platform, we consider CoFeB with PMA, characterized by a considerably larger $M_\mathrm{s}$ and stronger anisotropy, to explore a qualitatively different spectral regime. Specifically, the uniform-mode resonance frequency in CoFeB exceeds the applied microwave frequency, rendering excitation at the fundamental frequency subresonant. Higher-frequency components generated by the SC antenna, however, can couple to propagating spin-wave modes, enabling their selective excitation.

The Ga:YIG film has a thickness $d$ of $20$\,nm; in the simulations, it is modeled by a $0.765\,\mu$m-wide computational cell with periodic boundary conditions along the antenna axis. Its PMA stabilizes an out-of-plane magnetized state, while an externally applied out-of-plane bias field of $B_0=30$\,mT fixes the magnetization direction. The SC strip is assumed to be made of lead (Pb), with a width $w=500$~nm and a thickness $d_\mathrm{SC}=50$ nm. The SC antenna and Ga:YIG film are separated by a nonmagnetic insulating spacer of thickness $s=5$\,nm. At the applied magnetic field $B_0=30$\,mT, these dimensions ensure that the strip accommodates a single row of Abrikosov vortices in the mixed state, while providing sufficient space for their harmonic motion, as discussed in \textcolor{blue}{Appendix~B}. The ac current in the SC antenna drives the vortices into oscillatory motion via the time-dependent Lorentz force. 
As discussed in Sec.~\ref{sec:res}, the resulting vortex motion and current redistribution generate higher harmonics in the dynamic magnetic field of the SC antenna, see Fig.~\ref{fig:system}.

The choice of Pb as the SC antenna material is motivated by its relatively small London penetration depth (see Table~\ref{tab:tdgl_material_p}), which gives rise to strong, spatially localized vortex stray fields with large field gradients. These gradients enhance the vortex-induced redistribution of the SC transport current, thereby promoting the generation of higher harmonics in the antenna near field. For comparison, Table~\ref{tab:tdgl_material_p} also includes NbC as a dirty-limit SC with a Ginzburg–Landau parameter $\kappa \sim 100$, representing the opposite limit of strongly type-II behavior relative to Pb with $\kappa \approx 2.2$. Together, Pb and NbC span a broad range of Ginzburg–Landau parameters, encompassing values characteristic of many technologically relevant SCs, including Nb ($\kappa \approx 7$) \cite{Gol19adm,Dob19nph} and YBCO (in-plane $\kappa \approx 100$) \cite{Ghi26mmm}. Notably, NbC has previously been employed to demonstrate spin-wave generation via a distinct, Cherenkov-like mechanism \cite{Bes14prb,Phi23mmm,Dob25nan}. However, its comparatively weak vortex stray-field gradients make it less favorable for the higher-harmonic spin-wave generation mechanism investigated here. As our quantitative analysis below suggests, relatively clean Nb \cite{Dob12tsf} and potentially MgB$_2$ (effective $\kappa \sim 20$) \cite{Sch26prb} may be further alternative SC materials for realizing the effects discussed in this work.

\begin{figure}[t!]
\centering
\includegraphics[width=8.8cm]{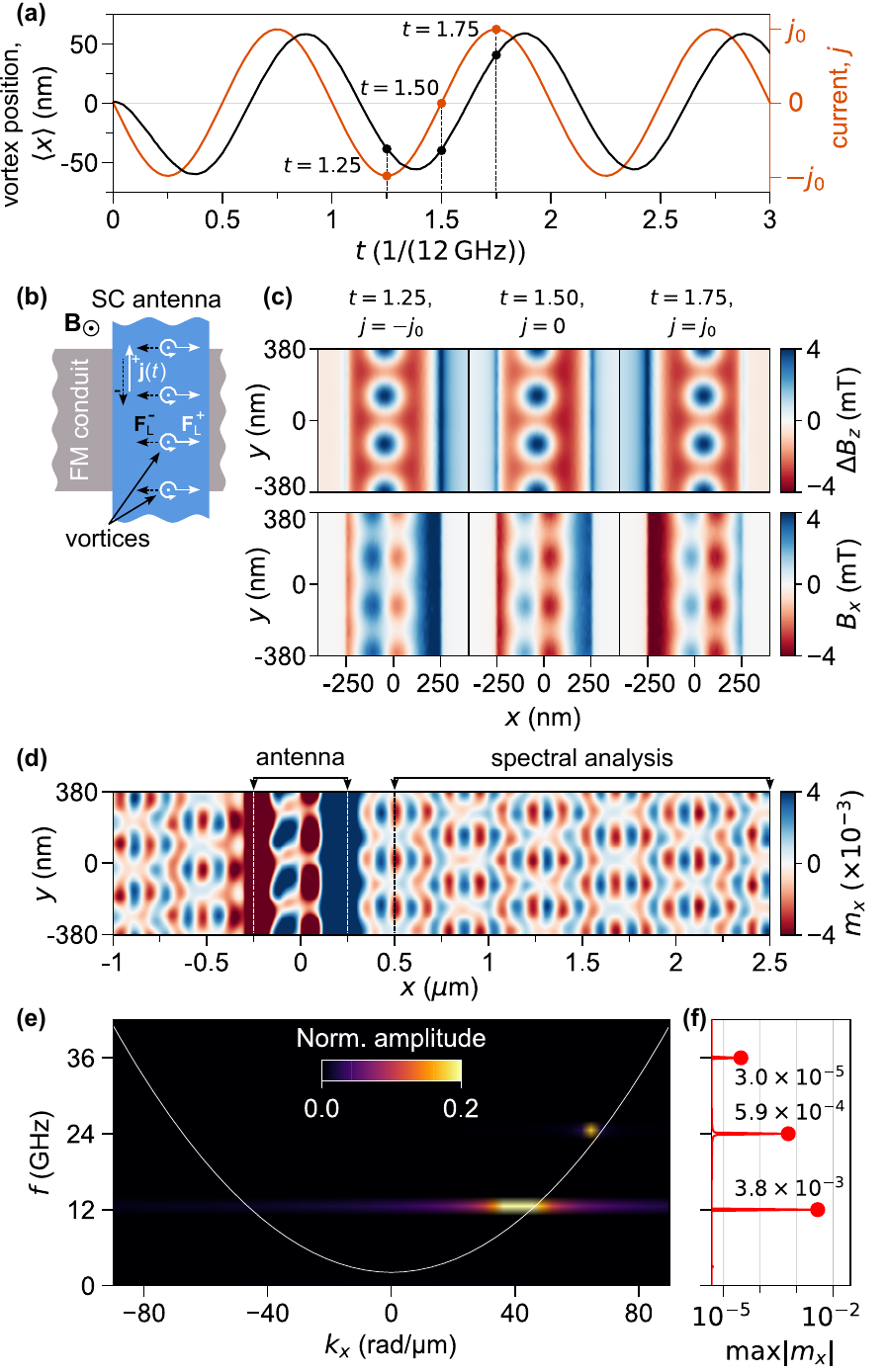}
\caption{
(a) Vortex positions under an ac current density of amplitude $j_0=0.032\,j_{\mathrm{TDGL}}$, where $j_{\mathrm{TDGL}}$ is the TDGL current-density unit (\textcolor{blue}{Appendix~D}).
(b) Schematic of the SC antenna above the FM conduit. The ac-current-induced Lorentz force $F_\mathrm{L}$ drives vortex oscillations about their equilibrium positions.
(c) In-plane $B_x$ and out-of-plane $\Delta B_z$ field distributions at the three instants marked in (a), calculated 5 nm below the SC antenna at $B_0=30$ mT, with $B_z=B_0+\Delta B_z$.
(d) Steady-state spatial distribution of $m_x$ (in units of $M_\mathrm{s}$) in the Ga:YIG conduit magnetized out of plane at $B_0=30$ mT. Vertical dashed lines mark the SC antenna edges, 5 nm above the film, and the region used for spectral analysis.
(e) Frequency–wave-vector Fourier amplitude spectrum of the space–time magnetization within the region marked in (d), averaged along the periodic $y$ direction to retain only $k_y=0$ (spin waves propagating along $x$). The linear color scale is normalized to the maximum amplitude and saturated at 20\%. Enhanced intensity at 12 and 24 GHz indicates fundamental and second-harmonic excitation, respectively. The white line denotes the analytical spin-wave dispersion.
(f) Maximum temporal Fourier amplitude within the region marked in (d).
}
\label{fig:fig2}
\end{figure}

The low-$M_\mathrm{s}$ Ga:YIG film and the FVSW geometry were chosen to suppress intrinsic nonlinear frequency conversion in the magnetic layer. At the excitation frequency $f$ of $12$\,GHz, well above the uniform-mode resonance at $2.93$\,GHz, the relevant FVSW mode has a finite wave number, $k\simeq43\,\mathrm{rad}\,\mu\mathrm{m}^{-1}$, see Fig.~\ref{fig:fig2}. Despite this relatively large wave vector, the low $M_\mathrm{s}$ keeps the dipolar contribution to the mode ellipticity weak, so that the magnetization precession remains nearly circular. At $12$\,GHz, the ratio of the two transverse dynamic magnetization amplitudes, $m_y$ and $m_x$, is $|m_y|/|m_x|\simeq1.008$, corresponding to an ellipticity parameter $|\epsilon|\simeq7.8\times10^{-3}$. As discussed in \textcolor{blue}{Appendix A}, such weak ellipticity strongly suppresses conventional second-harmonic generation associated with elliptical precession. The same material and geometrical conditions also weaken the coupling responsible for parallel parametric pumping, keeping the corresponding instability below threshold. Thus, both conventional nonlinear mechanisms considered here---ellipticity-induced second-harmonic generation and parallel parametric pumping---are suppressed under the present conditions. The exclusion arguments for nonlinear magnetic processes in Ga:YIG are further detailed in \textcolor{blue}{Appendix~A}.

Ga:YIG also has a very low Gilbert damping parameter $\alpha$ (see Table~\ref{tab:tdgl_material_p}). Although low damping enhances the resonant response and can lower the threshold for parametric instabilities, it is advantageous here because it allows the spin waves excited by the SC antenna to propagate over sufficiently long distances to be clearly resolved in the simulations. The choice of Ga:YIG therefore represents a compromise between suppressing intrinsic nonlinear magnetic effects and maintaining efficient spin-wave propagation.

Another advantage of the low-$M_{\rm s}$ Ga:YIG conduit is the suppression of magnetic back-action of the FM on the SC. The dynamic stray field generated by the magnetic texture or spin waves scales with the magnetization amplitude and is therefore strongly reduced in Ga:YIG \cite{Kha24pra}. For the Ga:YIG parameters considered here, our estimates show that this feedback field is several orders of magnitude smaller than the applied bias field. Its influence on the SC current and vortex dynamics can therefore be neglected, allowing the Pb/Ga:YIG system to be treated within a one-way coupling approximation. This is useful for the present purpose because it allows the higher-harmonic generation mechanism of the SC antenna to be isolated without introducing an additional self-consistent FM--SC feedback problem \cite{Yut26phr}.

For a high-$M_{\rm s}$ magnetic material, the same quantitative argument does not necessarily apply, and magnetic back-action on the SC can in principle become stronger. A fully self-consistent description would then require coupling the micromagnetic dynamics of the magnetic layer to the time-dependent Ginzburg--Landau dynamics within SC. Such a coupled FM--SC problem is of interest in its own right but lies beyond the scope of the present study. The principal demonstration of the higher-harmonic generation mechanism is therefore based on the Pb/Ga:YIG system, for which the one-way coupling approximation is well justified.

\begin{table}[t!]
\caption{Material parameters used in the simulations. The parameter sets are chosen to represent realistic material properties but do not necessarily reproduce the complete sets of values reported in the respective references.}
\label{tab:tdgl_material_p}
\small
\begin{center}
\begin{tabular}{lccc}
\hline
\multicolumn{4}{c}{\textbf{Superconducting materials}} \\
\hline
\textbf{Parameter} & \textbf{Quantity} & \textbf{Pb}\,\cite{Emb17nac} & \textbf{NbC}\,\cite{Dob20nac} \\
\hline
Clean coher. len., nm
& $\xi_0$
& 70
& 20 \\
Electron mfp, nm
& $l$
& 40
& 2 \\
Coherence len., nm
& $\xi(0)=0.85\sqrt{\xi_0l}$
& 45
& 5 \\
London pen. dep., nm
& $\lambda(0)$
& 100
& 500 \\
Strip thickness, nm
& $d_\mathrm{SC}$
& 50
& 50 \\
Stripe width, nm
& $w$
& 500
& 500 \\
SC--FM separation, nm
& $s$
& 5
& 5 \\
Inelastic scatt. coeff.
& $\Gamma$
& 5
& 30 \\
Critical temperat., K
& $T_\mathrm{c}$
& 7.2
& 5.6 \\
Upper critical field, T
& $B_{\mathrm{c}2}(0)$
& 0.15
& 12.5 \\
% Diffusion coeff., cm$^2$/s
% & $D$
% & 50
% & 0.5 \\
Normal resist., $\mu\Omega\cdot\mathrm{cm}$
& $\rho$
& 2
& 150 \\
Reduced temperature
& $T/T_\mathrm{c}$
& 0.5
& 0.8 \\
Coher. length at $T$, nm
& $\xi(T)$
& 64
& 6 \\
Penetr. depth at $T$, nm
& $\lambda(T)$
& 115
& 1100 \\
\hline\\[-1.2ex]\hline
\multicolumn{4}{c}{\textbf{Magnonic materials}} \\
\hline
\textbf{Parameter} & \textbf{Symbol} & \textbf{Ga:YIG} \cite{Bot22apl} & \textbf{CoFeB} \cite{Con13jap} \\
\hline
Film thickness, nm
& $d$
& 20
& 1 \\
Satur. magnetiz., kA/m
& $M_\mathrm{s}$
& 16
& 1200 \\
Exchange stiffn., pJ/m
& $A_\mathrm{ex}$
& 1.37
& 15 \\
PMA, J/m$^3$
& $K_\mathrm{u}$
& 756
& $1.176\!\times\!10^{6}$ \\
Reduced anisotropy
& $Q=\dfrac{K_\mathrm{u}}{\frac{1}{2}\mu_0 M_\mathrm{s}^2}$
& 4.70
& 1.30 \\
Gyrom. rat., rad/(T\,ns)
& $\gamma$
& 179
& 176 \\
Gilbert damping
& $\alpha$
& $10^{-4}$
& 0.006 \\
Static extern. field, mT
& $B_0$
& 30
& 30 \\
\hline
\end{tabular}
\end{center}
\end{table}

Accordingly, a 1-nm-thick CoFeB film with strong PMA serves as a complementary magnonic conduit. The pronounced PMA stabilizes an out-of-plane magnetized state supporting FVSW propagation despite the large $M_\mathrm{s}$. The film thickness and interfacial PMA are chosen such that the uniform-mode resonance frequency of $13.52$\,GHz exceeds the applied microwave frequency $f$. The excitation is therefore sub-resonant with respect to the lowest spin-wave mode and cannot resonantly build up a large precession amplitude. At the same time, higher-frequency components generated by the SC antenna can access propagating spin-wave states. This makes CoFeB particularly useful as a complementary test of higher-harmonic excitation in a spectral regime in which the response at the fundamental frequency is suppressed.

The FVSW geometry further reduces conventional nonlinear magnetic effects: the low-$k$ precession remains nearly circular, which weakens ellipticity-mediated nonlinear coupling, while the first-order parallel-pumping instability is suppressed because spin-wave states at half the pump frequency are absent below the bottom of the spin-wave spectrum. Thus, CoFeB remains a comparatively weakly nonlinear magnetic receiver under the conditions considered here despite its substantially larger $M_\mathrm{s}$. Finally, unlike YIG-based films, which are commonly grown on paramagnetic GGG substrates whose low-temperature magnetization can broaden the magnon linewidth and perturb the spin-wave spectrum, CoFeB can also be deposited on nonmagnetic substrates and therefore avoids this substrate-induced contribution \cite{Kos19apl,Ser25mtq}.

\section{Results\label{sec:res}}
The central aim of this study is to identify which spin-wave modes in the magnonic conduit can be excited by the vortex-containing SC antenna. Our analysis is based on numerical simulations of the SC and magnetic subsystems. The microwave field generated by the current-driven SC antenna is obtained from the TDGL equation and subsequently used as the excitation field in micromagnetic simulations of the spin-wave dynamics. We focus on whether vortex motion and the associated current redistribution generate higher-harmonic microwave fields that can excite spin waves beyond the drive frequency. The spectral content of the SC antenna field is analyzed in \textcolor{blue}{Appendix~C}, while details of the numerical implementation are provided in \textcolor{blue}{Appendix~D}. The modeling results are discussed next.

\begin{figure*}[]
\centering
\includegraphics[width=18cm]{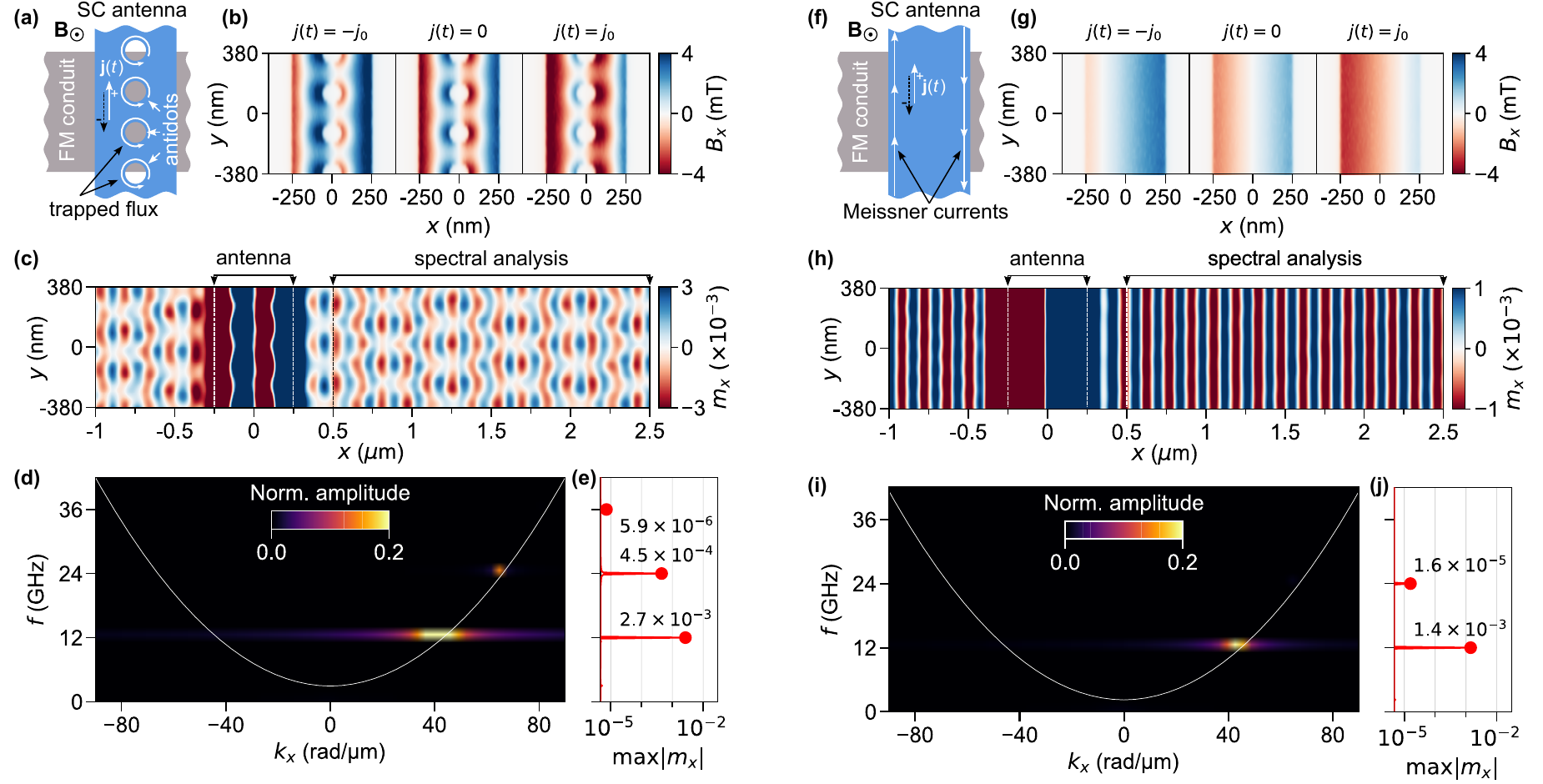}
\caption{Comparison of spin-wave excitation in a Ga:YIG conduit driven by an SC antenna with antidot-pinned vortices (a--e) and in the vortex-free Meissner state (f--j).
(a,f) Schematics of the two configurations.
(b,g) Spatial distributions of the in-plane rf field $B_x$ at $j=\pm j_0$ and $j=0$.
(c,h) Steady-state spatial distributions of $m_x$, indicating the antenna edges and spectral-analysis regions.
(d,i) Corresponding frequency--wave-vector Fourier amplitude spectra.
(e,j) Maximum temporal Fourier amplitudes.
The pinned-vortex state exhibits pronounced higher harmonics, whereas the Meissner-state response is dominated by the fundamental at $f=12$~GHz, with a weak residual second harmonic at $2f$.
Plotting conventions and analysis details are as in Fig.~\ref{fig:fig2}.
}
\label{fig:fig3}
\end{figure*}

\subsection{Higher-order harmonic spin-wave excitation mediated by the superconducting antenna}

Figure~\ref{fig:fig2}(a) shows the nearly harmonic ac-driven oscillations of the vortices about their equilibrium positions, as illustrated in Fig.~\ref{fig:fig2}(b). The corresponding out-of-plane magnetic-field distributions at three characteristic instants of the oscillation cycle are shown in Fig.~\ref{fig:fig2}(c). As the vortices move, their strongly localized stray fields shift relative to the magnetic layer and are accompanied by a redistribution of the screening and transport currents in the SC strip. Consequently, the local microwave field experienced by the magnetic layer acquires a non-sinusoidal time dependence despite the nearly harmonic vortex motion.

The resulting spin-wave excitation in Ga:YIG is illustrated in Fig.~\ref{fig:fig2}(d), which shows the steady-state spatial distribution of the dynamic magnetization component $m_x$: Spin waves emitted from the vicinity of the SC antenna propagate away from it. To identify their spectral content, we analyze the space--time magnetization dynamics in the $2\,\mu\mathrm{m}$-long region indicated by the dashed rectangle in Fig.~\ref{fig:fig2}(d), starting $250$\,nm from the right-hand edge of the antenna.

The frequency--wave-vector spectrum in Fig.~\ref{fig:fig2}(e) reveals two dominant propagating contributions at $12$ and $24$\,GHz. Both follow the analytical spin-wave dispersion, demonstrating that they correspond to propagating spin-wave modes rather than to a local response confined to the antenna region. The $12$\,GHz component is excited directly at the drive frequency $f$, whereas the $24$\,GHz branch represents excitation at the second harmonic, $2f$. The Fourier analysis in Fig.~\ref{fig:fig2}(f) confirms this result: the maximum amplitudes within the analysis region, which include obliquely propagating waves, at $12$ and $24$\,GHz are $3.8\times10^{-3}$ and $5.9\times10^{-4}$, respectively, corresponding to a second-harmonic amplitude of about $16\%$ of the fundamental. A weaker propagating contribution is also resolved at the third harmonic, $3f=36$\,GHz, with an amplitude of $3\times10^{-5}$, about $0.8\%$ of the fundamental.

The simultaneous appearance of a pronounced $2f$ signal and weaker higher harmonics despite the nearly harmonic vortex motion and essentially linear magnetization dynamics, with magnon--magnon interactions strongly suppressed, is a central result.  Efficient excitation of a given spin-wave mode requires the excitation field to contain both a frequency component matching the mode frequency and a spatial Fourier component compatible with its wave vector. This analysis therefore allows us to identify the excited modes in the frequency--wave-vector domain and to assess the role of the spatial structure of the excitation field. 
% In our system, the SC antenna with a row of vortices generates a strongly nonuniform current and stray-field distribution, resulting in a broad wave-vector spectrum. This enables efficient excitation of short-wavelength spin waves associated with the higher harmonics. 
%\textcolor{magenta}{In our system, the in-plane field $B_x$ above the vortex row oscillates nearly in antiphase with that near the strip edges [Fig.~\ref{fig:fig2}(c)]. This sign alternation across the strip enhances the spatial Fourier components of the field at wavelengths comparable to the strip width, favoring the excitation of such spin waves, while the periodicity of the vortex row along $y$ provides components with $k_y=\pm2\pi/a_\mathrm{v}$, where $a_\mathrm{v}$ is the vortex spacing, which launch obliquely propagating spin waves and cause the $y$-modulation of $m_x$ in Fig.~\ref{fig:fig2}(d).}
In the presence of vortices, the microwave in-plane field component $B_x$ beneath the SC strip [Fig.~\ref{fig:fig2}(c)] exhibits an additional phase reversal across the strip width (along the $x$ direction) compared with the vortex-free Meissner state [Fig.~\ref{fig:fig3}(f)]. This modified transverse field profile changes the wave-vector content of the antenna field -- see Fig.~\ref{fig:SM_antenna_spectrum}(a) in Appendix~C. Additionally, the periodic vortex arrangement along the strip axis $y$ modulates $B_x$ with the vortex spacing $a_\mathrm{v}$, introducing Fourier components at $k_y=\pm 2\pi/a_\mathrm{v}$, as presented in Fig.~\ref{fig:SM_antenna_spectrum}(b,c) in \textcolor{blue}{Appendix~C}. These components excite obliquely propagating spin waves with opposite $k_y$, consistent with the observed $y$-modulation of the corresponding phase-front pattern in Fig.~\ref{fig:fig2}(d).
%In a conventional stripline antenna, accessing comparable wave vectors would require a substantially narrower strip, which would compromise impedance matching and increase the current density and associated Joule heating.

%The dominance of the second harmonic can be anticipated from the spatial symmetry of the stray field generated by the oscillating vortex and the accompanying SC current distribution. In the present geometry, the in-plane field changes sign twice during one vortex oscillation cycle and therefore follows the fundamental frequency $f$. In contrast, the out-of-plane field at a given location can exhibit two pulses of the same sign during one vortex cycle, resulting in a strong rectification effect and a dominant $2f$ component -- see Fig.~\ref{fig:dipole}(c). The full spatial structure of both field components, however, gives rise to additional higher harmonics. Thus, the frequency conversion is not restricted to a single second-harmonic process but results from the spatial-to-temporal transduction of the strongly nonuniform magnetic field by the oscillating vortex configuration -- see discussion in Sec.\ref{sec:disc}B.}

To isolate the role of Abrikosov vortices in the harmonic generation, we compare the spin-wave response of the SC antenna with two reference cases: a mixed state with vortices pinned by an antidot lattice and the vortex-free Meissner state. The antidots were introduced by arranging a row of holes with a radius of $\xi$ along the $y$-direction at the center of the SC antenna, with a periodicity of 255 nm, matching the vortex spacing $a_\mathrm{v}$, see Fig.~\ref{fig:fig3}(a,b). In the second reference simulation, the SC antenna in the Meissner state was considered under a weak out-of-plane magnetic field of $B_0 = 5$ mT, see Fig.~\ref{fig:fig3}(f,g). Figures~\ref{fig:fig3}(c,h) show the corresponding steady-state distributions of $m_x$ in the Ga:YIG conduit. In both cases, the antenna excites propagating spin waves at the drive frequency, but the spectral content differs markedly. With pinned vortices, the $m_x$ map [Fig.~\ref{fig:fig3}(c)] shows a $y$-modulation similar to that in Fig.~\ref{fig:fig2}(d), originating from the periodic row of vortices and antidots, whereas in the Meissner state [Fig.~\ref{fig:fig3}(h)] the wavefronts are straight because the antenna field is uniform along $y$.

For the pinned-vortex state, the frequency--wave-vector spectrum in Fig.~\ref{fig:fig3}(d) contains pronounced signals at both $12$ and $24$~GHz, following the analytical spin-wave dispersion. The corresponding frequency spectrum in Fig.~\ref{fig:fig3}(e) exhibits amplitudes of $2.66\times10^{-3}$ and $0.45\times10^{-3}$ at $f$ and $2f$, respectively, so that the second-harmonic amplitude reaches about $17\%$ of the fundamental. A weaker third-harmonic signal at $3f=36$~GHz is also resolved, with an amplitude of $5.9\times10^{-6}$, about $0.22\%$ of the fundamental. Thus, higher-harmonic generation persists even when the vortices are confined, since pinned vortices still oscillate, but with a slightly smaller amplitude \cite{Dob15apl,Pom08prb}. Experimentally, such confinement of ac-driven vortices can be achieved using an artificial pinning landscape \cite{Wor12prb}.

By contrast, in the vortex-free Meissner state [Figs.~\ref{fig:fig3}(f)--\ref{fig:fig3}(j)], only the fundamental response at $12$\,GHz is prominent. The frequency spectrum in Fig.~\ref{fig:fig3}(j) exhibits amplitudes of $1.4\times10^{-3}$ and $1.6\times10^{-5}$ at $f$ and $2f$, respectively, so that the second-harmonic amplitude is only about $1\%$ of the fundamental, while no third-harmonic signal is resolved. 
This comparison demonstrates that pronounced higher-harmonic spin-wave excitation occurs in the vortex-containing SC state, whereas it is strongly suppressed in the vortex-free Meissner state under the respective magnetic-field conditions. Together with Fig.~\ref{fig:fig2}, these results identify the vortex dynamics and associated redistribution of SC currents as the source of the nonlinear microwave field responsible for the higher-harmonic spin-wave excitation.

\subsection{Spin-wave excitation in the absence of the fundamental mode}

\begin{figure}[t!]
\centering
\includegraphics[width=8.8cm]{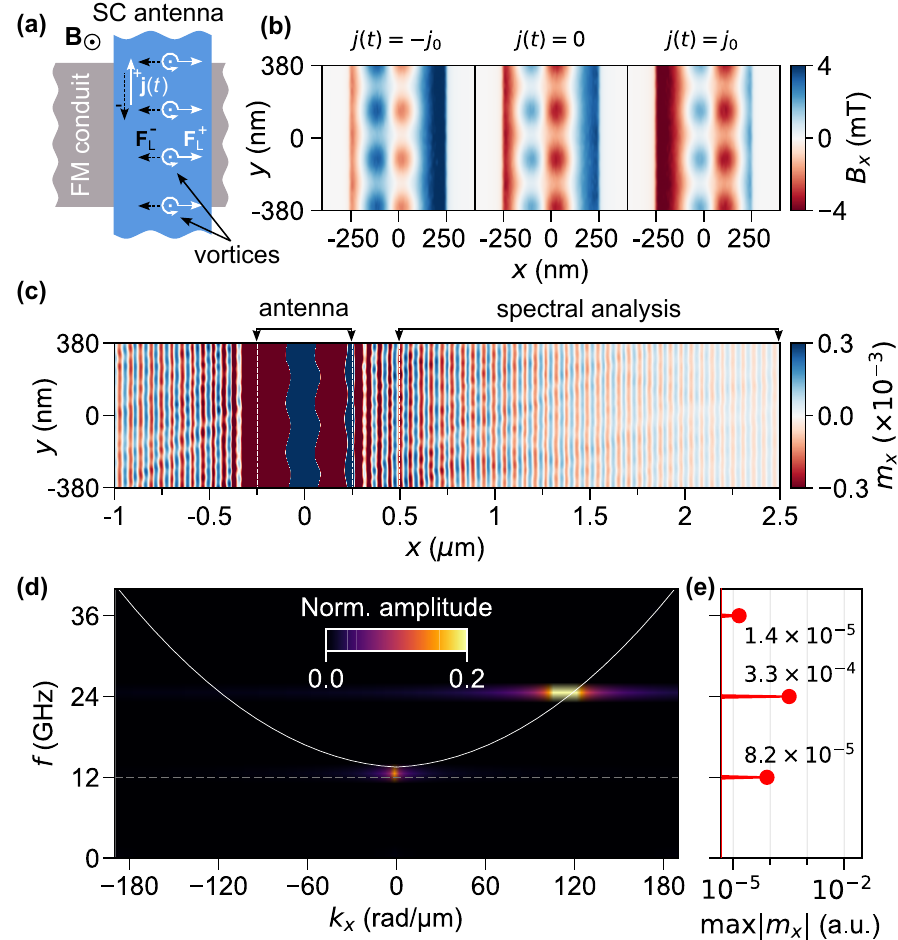}
\caption{Second-harmonic spin-wave generation in a CoFeB conduit driven by an SC antenna with unpinned vortices. The fundamental at $f=12\,\mathrm{GHz}$ lies below the FMR frequency and cannot excite propagating spin waves.
(a) Geometry.
(b) Spatial distributions of $B_x$ at $j=\pm j_0$ and $j=0$.
(c) Steady-state $m_x$ distribution and (d) corresponding frequency--wave-vector Fourier amplitude spectrum. The dashed line marks the fundamental at $f$, while the second harmonic at $2f=24\,\mathrm{GHz}$ excites propagating spin waves.
(e) Maximum temporal Fourier amplitudes.
Plotting conventions and analysis details are as in Fig.~\ref{fig:fig2}.
}
\label{fig:fig4}
\end{figure}

We next consider a regime in which the drive frequency lies below the bottom of the propagating spin-wave spectrum. The geometry in Fig.~\ref{fig:fig4}(a,b) is the same as in Fig.~\ref{fig:fig2}(b,c) and Fig.~\ref{fig:fig4}(c-e) shows the response of the CoFeB conduit for a drive at $f=12$\,GHz, while the uniform-mode resonance is at $13.52$\,GHz. Consequently, the fundamental excitation is sub-resonant and cannot efficiently launch propagating spin waves. The steady-state $m_x$ distribution in Fig.~\ref{fig:fig4}(c) nevertheless reveals a wave propagating away from the SC antenna.

The frequency--wave-vector spectrum in Fig.~\ref{fig:fig4}(d) identifies this excitation as the second harmonic at $2f=24$\,GHz. While the $12$\,GHz component lies below the spin-wave band and remains nonpropagating, the $24$\,GHz component intersects the spin-wave dispersion and therefore excites a propagating mode. The temporal Fourier amplitudes in Fig.~\ref{fig:fig4}(e) confirm this behavior: the response at $12$\,GHz is only $8.2\times10^{-5}$, whereas the second-harmonic peak reaches $3.3\times10^{-4}$.

This regime provides a particularly clear demonstration of frequency conversion by the SC antenna. Propagating spin waves appear even though the fundamental microwave frequency cannot excite the magnetic conduit directly. The excitation is therefore enabled by the higher-harmonic component generated by the vortex-containing SC antenna, which shifts spectral weight into the propagating spin-wave band.

\section{Discussion}\label{sec:disc}
\subsection{Higher-harmonic spin-wave excitation}

The simulations establish three main results. First, an SC antenna in the mixed state generates a pronounced second-harmonic microwave field and excites spin waves at $2f$, even though the ac-driven vortex displacement remains nearly harmonic in time. Second, the $2f$ spin-wave response persists for both freely oscillating and pinned vortices, whereas it is strongly suppressed in the vortex-free Meissner state, identifying the vortex-containing SC state as the origin of the enhanced frequency conversion. Third, in the CoFeB conduit, where the fundamental drive frequency lies below the bottom of the propagating spin-wave spectrum, the generated second harmonic nevertheless excites a propagating mode at $2f$. Together with the suppression of conventional nonlinear processes in the magnetic layer outlined above and detailed in \textcolor{blue}{Appendix A}, these results show that the magnetic conduit primarily acts as a receiver of the nonlinear microwave field generated by the SC antenna.

Our results suggest that the higher-harmonic response does not require strongly anharmonic vortex motion. Instead, it originates from the conversion of a nearly harmonic vortex displacement into a nonsinusoidal local electromagnetic response. As a vortex oscillates about its equilibrium position, its strongly nonuniform stray-field profile moves relative to a fixed observation point, while the vortex motion simultaneously redistributes the transport and screening currents. The resulting local magnetic field therefore need not follow the nearly sinusoidal time dependence of the vortex displacement and can acquire higher temporal harmonics.

The pronounced second harmonic can be understood from the spatial structure of this electromagnetic field. During one oscillation period, a vortex can approach a given observation region \emph{twice}, producing two field variations within a single drive cycle and thereby enhancing the component at $2f$. Away from high-symmetry positions, the two parts of the oscillation need not produce identical local fields because of the spatially nonuniform vortex field and current distribution. Consequently, additional higher harmonics can also arise. Their relative amplitudes are expected to depend on the observation position, vortex-oscillation amplitude, and spatial profiles of the vortex stray field and redistributed currents.

This mechanism differs from conventional sources of nonlinear microwave response in SCs, where higher-harmonic generation is commonly associated with trapped flux, nonlinear current--voltage characteristics, or strongly anharmonic vortex dynamics \cite{Ji89,Shk08prb,Checchin20}. Here, by contrast, the vortex displacement remains close to sinusoidal while a pronounced $2f$ component develops in the local magnetic field. The relevant nonlinearity therefore resides primarily in the spatial dependence of the electromagnetic response sampled by the oscillating vortices and in the accompanying current redistribution, rather than in a strongly anharmonic time dependence of the vortex displacement itself.

The effect can thus be viewed as a \emph{spatial-to-temporal transduction}: nearly harmonic motion through a strongly nonuniform electromagnetic field converts spatial variation into temporal harmonics at a fixed observation point. Related concepts occur in cavity optomechanics, where harmonic mechanical motion is transduced through a nonlinear position dependence of the optical response \cite{Leijssen17}, and in dynamic atomic force microscopy, where an oscillating probe samples a nonlinear tip--sample interaction and thereby generates higher harmonics \cite{Stark03,Kuchuk15}. In the present system, the spatially varying response is provided by the vortex stray field together with the accompanying redistribution of SC currents.

An important implication of our studies for SC-FM magnonics is that the frequency conversion occurs directly in the magnetic near field of the antenna. A monochromatic drive at frequency $f$ can therefore generate field components at $2f$ and higher harmonics capable of resonantly exciting spin-wave modes inaccessible at the fundamental frequency. The CoFeB results provide a particularly direct demonstration: although $f=12$\,GHz lies below the propagating spin-wave spectrum, the generated $2f=24$\,GHz component launches a propagating spin wave. The SC antenna can therefore extend the accessible spin-wave frequency range without requiring nonlinear magnon dynamics or direct microwave excitation at the target frequency. 
% Moreover, achieving comparable spin-wave wave vectors with a conventional stripline antenna would require a substantially narrower conductor, posing challenges for impedance matching while increasing the current density and, consequently, resistive heating.
In addition, the vortex-induced nonuniformity of the antenna field across the strip (Sec.~\ref{sec:res} and Appendix~C) substantially enhances the excitation of spin waves with wavelengths between about half and the full strip width, which with a conventional stripline antenna would require a considerably  narrower conductor, posing challenges for impedance matching while increasing the current density and, consequently, resistive heating. The periodicity of the vortex row along the strip further enables the excitation of obliquely propagating spin waves, which a uniform stripline cannot launch.

\subsection{Mechanism of nonlinear spatial-to-temporal transduction}

The results above indicate that the predicted higher-harmonic spin-wave excitation originates predominantly from the vortex-containing SC antenna rather than from intrinsic nonlinearities of the magnetic layer. We now examine the physical origin of this frequency conversion and show how nearly harmonic vortex motion can generate a strongly nonsinusoidal local microwave field.

The higher-harmonic response does not require strongly anharmonic vortex motion or an intrinsically nonlinear magnetic response. Instead, a nearly harmonic displacement of a spatially nonuniform vortex-generated near-field profile relative to a fixed observation point produces a nonsinusoidal local time dependence---a mechanism we term \emph{spatial-to-temporal transduction}.

Let $X_v(t)$ denote the mean transverse coordinate of the vortex row. In the regime considered here,
\begin{equation}
    X_v(t)=X_0+a\cos(\omega t+\phi),
    \label{eq:vortex_motion}
\end{equation}
where $X_0$, $a$, $\omega=2\pi f$, and $\phi$ are the equilibrium position, oscillation amplitude, angular frequency, and phase relative to the microwave drive, respectively. In the harmonic approximation, $a = X_1$, where $X_1$ is the fundamental Fourier amplitude defined in \textcolor{blue}{Appendix B}. The vortex motion modifies the near field both through displacement of the localized vortex stray field and through redistribution of the transport and screening currents. The latter represents a periodic deformation of the current pattern rather than its rigid translation.

The local magnetic field may therefore be written as
\begin{equation}
    \mathbf{B}(\mathbf{r},t)
    =
    \mathbf{B}\!\left[\mathbf{r};I(t),X_v(t)\right],
    \label{eq:field_general}
\end{equation}
with the harmonic drive
\begin{equation}
    I(t)=I_0\cos\omega t.
    \label{eq:drive_current}
\end{equation}
Although both $I(t)$ and $X_v(t)$ can remain harmonic, the field at a fixed point need not depend linearly on them. Defining $\delta I=I(t)$ and $\delta X_v=X_v(t)-X_0$, a selected field component can be expanded around the equilibrium configuration  as
\begin{equation}
\begin{split}
    \delta B={}&
    B_I\,\delta I+B_X\,\delta X_v
    +\frac{1}{2}B_{XX}\,\delta X_v^2
    \\
    &+B_{IX}\,\delta I\,\delta X_v
    +\frac{1}{6}B_{XXX}\,\delta X_v^3+\cdots ,
\end{split}
\label{eq:field_expansion}
\end{equation}
where $B_I$, $B_X$, $B_{XX}$, $B_{IX}$, and $B_{XXX}$ denote the corresponding partial derivatives of the local field evaluated at equilibrium. Terms nonlinear in $\delta I$ alone are omitted from Eq.~\eqref{eq:field_expansion} to isolate the vortex-mediated spatial-to-temporal conversion mechanism investigated here from the intrinsic nonlinear electrodynamic response of the SC. Since $\delta I$ and $\delta X_v$ oscillate at $\omega$, the quadratic terms contain dc and $2\omega$ components, while the cubic term also generates a $3\omega$ component. Thus, higher temporal harmonics can arise even for nearly sinusoidal current and vortex displacement.

The geometrical origin of this effect becomes particularly evident when the localized vortex stray-field profile is approximated as translating rigidly with the vortex,
\begin{equation}
    B(x,t)=B_v\!\left[x-X_v(t)\right],
    \label{eq:translated_profile}
\end{equation}
where $x$ is the observation coordinate and $B_v(x)$ is the static spatial profile of the selected vortex-field component. With $s=x-X_0$ and $\Omega=\omega t+\phi$,
\begin{equation}
    B(x,t)=B_v\!\left[s-a\cos\Omega\right].
    \label{eq:translated_profile_xi}
\end{equation}
For $a$ small compared with the characteristic spatial scale of $B_v$, expansion in $a$ gives
\begin{equation}
\begin{split}
    B(x,t)={}&B_v(s)-aB_v'(s)\cos\Omega
    +\frac{a^2}{2}B_v''(s)\cos^2\Omega
    \\
    &-\frac{a^3}{6}B_v'''(s)\cos^3\Omega+\cdots .
\end{split}
\label{eq:profile_expansion}
\end{equation}
The leading second- and third-harmonic amplitudes are therefore
\begin{equation}
\begin{split}
    B_{2\omega}(s)&\simeq \frac{a^2}{4}B_v''(s),\\
    B_{3\omega}(s)&\simeq-\frac{a^3}{24}B_v'''(s).
\end{split}
\label{eq:harmonic_amplitudes}
\end{equation}
Hence, the second harmonic probes the local curvature of the spatial field profile, while the third harmonic is governed by its third spatial derivative. More generally, in the perturbative regime, higher temporal harmonics are controlled by progressively higher spatial derivatives of the moving near field.

\begin{figure}[t!]
    \centering
    \includegraphics[width=1\columnwidth]{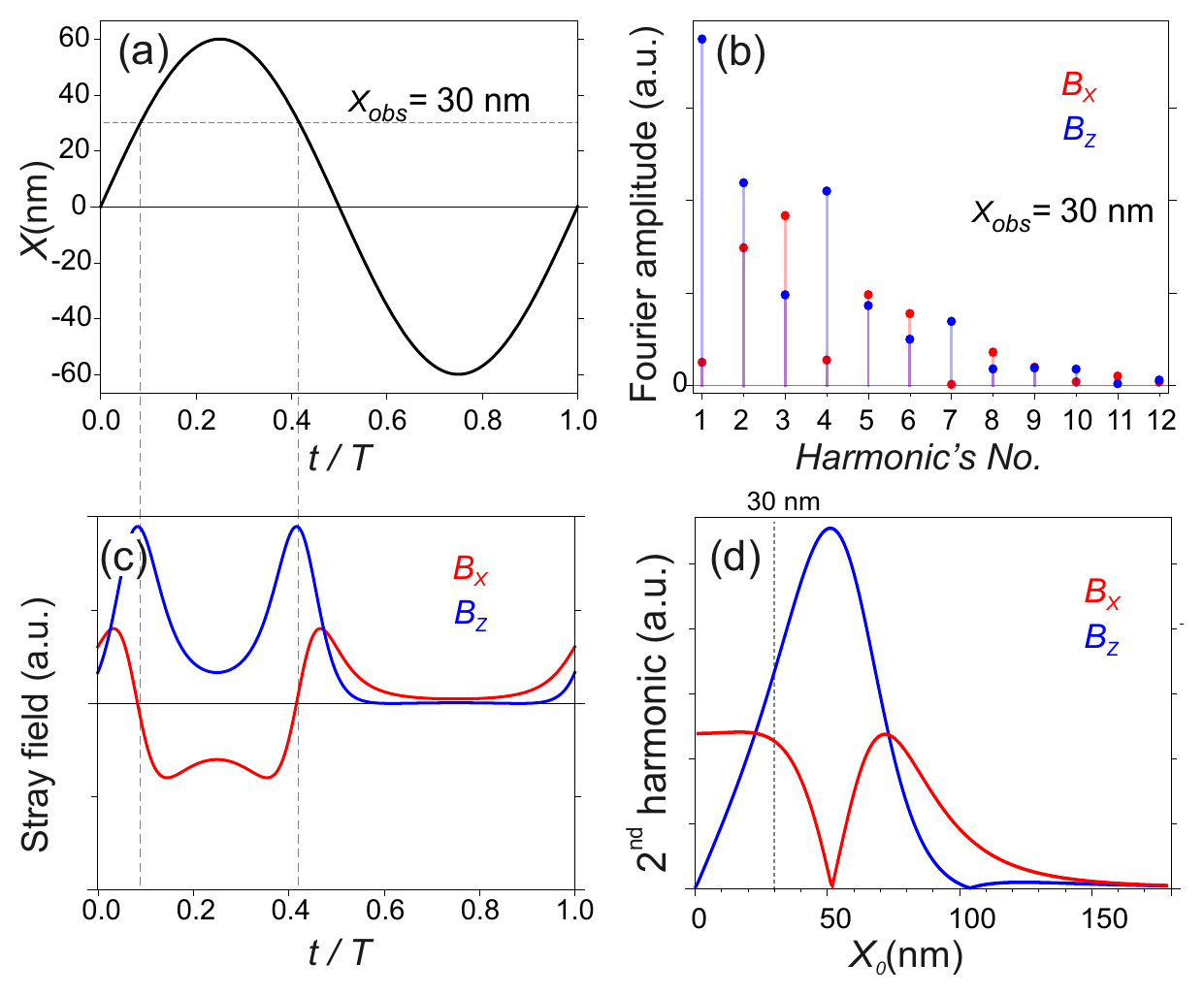}
\caption{Spatial-to-temporal transduction by an oscillating magnetic dipole.
(a) Harmonic displacement of a $z$-oriented dipole along $x$, with oscillation period $T$.
(b) Fourier amplitudes of the field components in (c), revealing higher harmonics of the driving frequency.
(c) Temporal evolution of the in-plane ($B_x$) and out-of-plane ($B_z$) field components at a fixed off-center observation point, $x_{\rm obs}=30$~nm. Despite purely harmonic dipole motion, the local field is strongly nonsinusoidal.
(d) Spatial dependence of the Fourier coefficients, highlighting the localized nature of harmonic generation.
}
    \label{fig:dipole}
\end{figure}

Figure~\ref{fig:dipole} provides a minimal illustration of this mechanism. The localized field profile is represented by the stray field of a $z$-oriented magnetic dipole undergoing harmonic motion, $X_v(t)=a\cos\omega t$, along $x$, corresponding to Eq.~\eqref{eq:vortex_motion} with $X_0=\phi=0$ for panels Fig.\,\ref{fig:dipole}(a–c). In panel Fig.\,\ref{fig:dipole}(d), $X_0$ is varied to illustrate the spatial dependence of the second-harmonic amplitude. The dipole is not intended as a microscopic model of an Abrikosov vortex; it represents a spatially localized field profile whose position varies periodically. For completeness, the field of the oscillating dipole can be written as
\begin{equation}
    \mathbf{B}_{\rm d}(\mathbf{r},t)
    =
    \frac{\mu_0}{4\pi}
    \left[
    \frac{3\bigl(\mathbf{m}\!\cdot\!\mathbf{R}\bigr)\mathbf{R}}{R^5}
    -
    \frac{\mathbf{m}}{R^3}
    \right],
    \label{eq:dipole_field}
\end{equation}
where $\mathbf m$ is the magnetic dipole moment and $\mathbf R=\mathbf r-\mathbf r_v(t)$ is the displacement vector from the instantaneous dipole position $\mathbf r_v(t)$ to the observation point $\mathbf r$.

Although its displacement is purely harmonic [Fig.~\ref{fig:dipole}(a)], the field measured at a fixed off-center point $x_{\rm obs}=30$~nm is strongly nonsinusoidal [Fig.~\ref{fig:dipole}(c)] and consequently contains higher harmonics [Fig.~\ref{fig:dipole}(b)]. Their amplitudes vary with the relative position of the dipole and the observation point [Fig.~\ref{fig:dipole}(d)], consistent with the dependence on the spatial derivatives in Eq.~\eqref{eq:harmonic_amplitudes}. In the FVSW geometry considered here, with the equilibrium magnetization along $z$, the in-plane component $B_x$ acts as a transverse excitation field and is therefore directly relevant for spin-wave excitation. In the actual SC antenna, the near field is more complex because vortex motion is accompanied by redistribution of the transport and screening currents; the dipole model serves only to isolate the geometrical origin of the frequency conversion.

% The effect is therefore intrinsically local.
The generation of harmonics in the near field is therefore local; spin-wave excitation, however, is governed by the wave-vector content of the entire antenna field [Eq.~\eqref{eq:linear_magnetic_response}].
If $L$ is the characteristic scale of the field variation, the $n$th spatial derivative scales roughly as $B_v/L^n$. In the perturbative regime, the relative strength of successive harmonics is therefore controlled by powers of the dimensionless ratio $a/L$. Higher-harmonic generation is weak for $a\ll L$ and increases as the oscillation amplitude becomes a larger fraction of the spatial variation scale. When $a$ becomes comparable to $L$, the truncated expansion in Eq.~\eqref{eq:profile_expansion} is no longer quantitatively valid; nevertheless, the full translated profile in Eq.~\eqref{eq:translated_profile_xi} can then produce a strongly nonsinusoidal waveform and substantial higher-harmonic content.

The resulting dynamics can be viewed as a two-step process. First, vortex motion and the accompanying redistribution of SC currents convert the monochromatic drive into a spatially localized near field containing $\omega$, $2\omega$, $3\omega,\ldots$. The magnetic layer then responds to these components. In the linear-response regime,
\begin{equation}
    m(n\omega,\mathbf{k})
    =
    \chi_m(n\omega,\mathbf{k})\,h(n\omega,\mathbf{k}),
    \label{eq:linear_magnetic_response}
\end{equation}
where $m(n\omega,\mathbf{k})$ is the spin-wave amplitude, $h(n\omega,\mathbf{k})= B(n\omega,\mathbf{k})/\mu_0$ the corresponding harmonic component of the microwave magnetic field expressed in A/m, $\chi_m(n\omega,\mathbf{k})$ the dynamic magnetic susceptibility, $\mathbf{k}$ the spin-wave wave vector, and $n=1,2,3,\ldots$ the harmonic order. Frequency conversion thus occurs predominantly in the SC antenna, while the magnonic conduit acts as a frequency- and wave-vector-selective receiver of the generated higher-harmonic near fields.

\subsection{Comparison with other fluxon-based mechanisms of microwave and magnon generation}

Finally, it is instructive to place the present mechanism of higher-harmonic spin-wave generation in the broader context of microwave and magnon emission by moving vortices. Coherently translating fluxons are known to generate a washboard
spectrum of microwave radiation \cite{Bul06prl}, with characteristic frequencies $f_n = nv/a_\mathrm{v}$, where $n=1,2,3,\ldots$ is the harmonic order, $v$ is the vortex velocity, and $a_\mathrm{v}$ is the vortex-lattice parameter \cite{Dob18nac}. This mechanism was predicted for vortex lattices reaching the edges of type-II SCs \cite{Bul06prl} and observed experimentally for SC-insulator multilayers \cite{Dob18nac}.

Another mechanism operates in Cherenkov-like magnon generation by fast dc-driven vortices \cite{Bes14prb,Phi23mmm}, where a traveling magnetic-field profile \cite{Dob18apl,Emb17nac} excites spin waves only when the vortex velocity exceeds the minimum phase velocity required by the velocity-dependent phase-matching condition. This Cherenkov-like emission is directional, with the excited spin waves propagating along the direction of vortex motion. This type of spin-wave generation was recently demonstrated experimentally for SC-FM Nb-C/Co-Fe heterostructures \cite{Dob25nan}. In both cases, the essential ingredient is the translational motion of a spatially nonuniform vortex stray field. 

The regime considered in this work is qualitatively different: the vortices remain \emph{confined} and undergo \emph{nearly harmonic ac oscillations} with essentially zero mean drift. Frequency conversion therefore requires neither washboard periodicity nor a Cherenkov-type critical vortex velocity. Instead, it arises from spatial-to-temporal transduction, whereby the oscillatory displacement of the strongly nonuniform vortex field, together with the accompanying redistribution of SC currents, generates higher temporal harmonics of the local microwave field. Their amplitudes evolve continuously with the vortex oscillation amplitude rather than appearing only above a velocity threshold. In the FVSW geometry considered here, the resulting second-harmonic spin waves propagate in \emph{both directions} away from the antenna, in contrast to the unidirectional Cherenkov-like emission. These harmonics can excite spin waves at $2f$ even when the fundamental frequency lies outside the propagating spin-wave band. 

In this way, vortex motion can generate microwave and magnon radiation through several distinct mechanisms such as washboard emission, thresholded directional Cherenkov-like excitation, and the confined ac-driven harmonic-generation mechanism demonstrated here.

\section{Conclusion}

To summarize, we have demonstrated that a superconducting microstrip antenna with a row of ac-driven vortices acts as a nonlinear microwave source for spin-wave excitation. Time-dependent Ginzburg--Landau and micromagnetic simulations show that vortex oscillations generate pronounced higher harmonics of the local microwave field even when the vortex displacement remains nearly harmonic. The resulting second harmonic excites propagating spin waves at $2f$ in both Ga:YIG and CoFeB conduits, whereas no comparable response occurs in the vortex-free Meissner state. In CoFeB, the fundamental drive lies below the propagating spin-wave spectrum, while the generated second harmonic couples to a propagating mode, providing a direct manifestation of frequency conversion by the superconducting antenna.
Compared with a vortex-free antenna of the same width and drive current, we predict substantially stronger excitation of spin waves with wavelengths between about half and the full strip width.

We explain the results of our simulations with a spatial-to-temporal transduction mechanism: harmonic motion of vortices through a strongly nonuniform near field, together with the associated redistribution of transport and screening currents, converts spatial field gradients into higher temporal harmonics. A simplified oscillating-dipole model captures the geometrical origin of this effect and shows that strongly nonsinusoidal local fields do not require intrinsically anharmonic source motion. Estimates of ellipticity-induced second-harmonic generation and parallel parametric pumping in Ga:YIG further show that conventional nonlinear magnetic processes are strongly suppressed under the conditions considered here. These results identify Abrikosov vortices as active nonlinear elements for local microwave frequency conversion and establish a route to extending the accessible spin-wave frequency range in superconducting--magnonic hybrids.

\begin{acknowledgments}
The work of A.P. was funded by the Deutsche Forschungsgemeinschaft (DFG, German Research Foundation) through Grant No.~583926508 (SyMPaThY) and by Germany's Excellence Strategy---EXC-2123 QuantumFrontiers---390837967, project Q-53 (Opto-Fluxonics). P.G., J.K., and J.W.K. acknowledge financial support from the National Science Centre, Poland, through Grant No.~UMO-2021/43/I/ST3/00550 and Grant No.~UMO-2023/49/B/ST3/02920. O.D. acknowledges financial support from the Austrian Science Fund (FWF) through Grant No.~I~6079 (FluMag) and from the DFG through Grant No.~577881064 (Super3DMag). A.P. gratefully acknowledges the use of the CryoCore and CryoCube simulation workstations at CryoQuant/TU Braunschweig and the support of the Braunschweig International
Graduate School of Metrology (B-IGSM). P.G. acknowledges the use of the infrastructure of the Poznań Supercomputing and Networking Center (PCSS) within Scientific Computing Space no. pl0804. This research is based on work from COST Actions CA21144 (SuperQuMap) and CA23134 (Polytopo), supported by the European Cooperation in Science and Technology.

\end{acknowledgments}

\section*{DATA AVAILABILITY STATEMENT}
The data supporting the findings of this study are openly
available  \cite{Pokusinskyi2026}. 

%\textcolor{cyan}{I prefer Zenodo. The paper will be used for reporting the grant and the publication of supporting dataset repository with DOI is obligatory. Zenodo is OK for NCN (there is some list of allowed repositories). I have to add grant acknowledgments to the grant in the description dataset. Anton please pass all you like to post (with proper description) to Julia. She will post it on Zenodo. }
%\textcolor{magenta}{Pawel: All data used to prepare the figures, as well as the micromagnetic simulation scripts, are ready to be uploaded to Zenodo, along with the time-dependent profiles used to excite SWs (maps $A_0$, $C_n$, and $S_n$ from Eq.~(\ref{eq:hrf}) and the comparison with Anton's data).}

\section*{Appendix A: NONLINEAR MAGNETIC PROCESSES}
\label{app:Nonlinear}

To distinguish harmonic generation by the SC antenna from intrinsic nonlinear magnetization dynamics, we estimate two nonlinear processes in Ga:YIG: ellipticity-induced second-harmonic generation and parallel parametric pumping. Throughout this Appendix, the transverse dynamic magnetization components $m_x$ and $m_y$, as well as $M_z$ and their corresponding amplitudes, are dimensional magnetization quantities expressed in A/m. Normalized amplitudes are written explicitly as ratios to the saturation magnetization $M_{\rm s}$. For the FVSW geometry, the equilibrium magnetization is oriented along $z$, while the spin-wave wave vector $\mathbf{k}$ is along $x$. The transverse dynamic components are represented as
$m_x(t)=\mathrm{Re}[\widetilde m_x e^{i\omega t}]$ and
$m_y(t)=\mathrm{Re}[\widetilde m_y e^{i\omega t}]$, with
$\widetilde m_x=m_{x0}$ and $\widetilde m_y=-i m_{y0}$, where $m_{x0}$ and $m_{y0}$ are the real positive amplitudes of the principal axes of the precession ellipse.
For the fundamental thickness mode,
$\omega_k=\gamma\sqrt{B_1B_2}$ and
$m_{y0}/m_{x0}=\sqrt{B_2/B_1}$
\cite{KalinikosSlavin86,StancilPrabhakar09}, with
$B_1=B_{\rm eff}+2A_{\rm ex}k^2/M_{\rm s}$,
$B_2=B_1+\mu_0M_{\rm s}P(kd)$, and
$P(kd)=1-(1-e^{-kd})/(kd)$.
Here, $B_{\rm eff}$ contains the external, anisotropy, and perpendicular demagnetizing fields; $A_{\rm ex}$ is the exchange stiffness, $M_{\rm s}$ is the saturation magnetization, $d$ is the magnetic-film thickness, $k=|\mathbf{k}|$, and $\gamma$ is the gyromagnetic ratio. The Gilbert relaxation rate is
$\Gamma_k\simeq\alpha\gamma(B_1+B_2)/2$, where $\alpha$ is the Gilbert damping parameter.

\textbf{Quadratic second-harmonic response}. 
Elliptical precession provides a conventional quadratic mechanism for second-harmonic generation \cite{Herrmann64}. For the Ga:YIG parameters used here, the $12$~GHz FVSW mode has
$k\simeq43~\mathrm{rad}\,\mu\mathrm{m}^{-1}$ and
$m_{y0}/m_{x0}\simeq1.008$, indicating nearly circular precession. The corresponding ellipticity,
\begin{equation}
\epsilon=
\frac{m_{y0}^2-m_{x0}^2}
     {m_{y0}^2+m_{x0}^2},
\label{eq:app_epsilon}
\end{equation}
is $|\epsilon|\simeq7.8\times10^{-3}$.

To second order in the transverse magnetization, conservation of the magnetization magnitude gives
\begin{equation}
M_z(t)\simeq
M_{\rm s}
-\frac{m_x^2(t)+m_y^2(t)}
       {2M_{\rm s}}.
\label{eq:app_Mz}
\end{equation}
The amplitude $m_{2\omega}$ of the resulting longitudinal component at $2\omega$ is therefore
\begin{equation}
\frac{m_{2\omega}}{M_{\rm s}}
=
\frac{|m_{y0}^2-m_{x0}^2|}
     {4M_{\rm s}^2}.
\label{eq:app_m2}
\end{equation}

For elliptical precession, we introduce the effective small-angle amplitude
\begin{equation}
\theta_{\rm eff}^2=
\frac{m_{x0}^2+m_{y0}^2}
     {2M_{\rm s}^2}.
\label{eq:app_theta}
\end{equation}
Unlike the cone angle of circular precession, $\theta_{\rm eff}$ is an amplitude parameter because the two principal precession amplitudes are slightly different. Equations~(\ref{eq:app_epsilon}) and (\ref{eq:app_theta}) give
\begin{equation}
\frac{m_{2\omega}}{M_{\rm s}}
=
\frac{|\epsilon|}{2}\theta_{\rm eff}^2
\simeq
3.9\times10^{-3}\theta_{\rm eff}^2.
\label{eq:app_m2theta}
\end{equation}

Equation~(\ref{eq:app_m2theta}) directly shows why this intrinsic nonlinear contribution is weak in the present Ga:YIG geometry. The second-harmonic response is quadratic in the precession amplitude and is additionally reduced by the small ellipticity factor $|\epsilon|/2\simeq3.9\times10^{-3}$. To make the suppression relative to the fundamental response explicit, we define the characteristic transverse amplitude
\begin{equation}
m_\perp=
\sqrt{\frac{m_{x0}^2+m_{y0}^2}{2}}
=
M_{\rm s}\theta_{\rm eff}.
\label{eq:app_mperp}
\end{equation}
The directly driven transverse response is linear in $\theta_{\rm eff}$, whereas the intrinsic second-harmonic response scales as $|\epsilon|\theta_{\rm eff}^2$. Their ratio is
\begin{equation}
\frac{m_{2\omega}}{m_\perp}
=
\frac{|\epsilon|}{2}\theta_{\rm eff}
\simeq
3.9\times10^{-3}\theta_{\rm eff}.
\label{eq:app_relative}
\end{equation}
Thus, the quadratic second-harmonic response is parametrically small: it contains two independent suppression factors, the small ellipticity $|\epsilon|\ll1$ and an additional power of the small precession amplitude $\theta_{\rm eff}\ll1$. Consequently, within the perturbative regime, the intrinsic second harmonic remains much smaller than the directly driven fundamental response.

\textbf{Parallel parametric pumping}. A second nonlinear channel is conventional parallel parametric pumping. In the perpendicular configuration, the out-of-plane microwave-field component is parallel to the equilibrium magnetization and can create magnon pairs with opposite wave vectors satisfying
\begin{equation}
\omega_{\mathbf{k}}+\omega_{-\mathbf{k}}=\omega_{\rm p},
\label{eq:app_pp_pair}
\end{equation}
where $\omega_{\rm p}$ is the pump angular frequency \cite{Braecher17,Kostylev95}. For reciprocal FVSWs, this reduces to
\begin{equation}
\omega_k=\frac{\omega_{\rm p}}{2},
\label{eq:app_pp_res}
\end{equation}
so a $12$~GHz pump can parametrically generate magnons at $6$~GHz.

For the present dispersion, the $6$~GHz mode has
$k\simeq24.7~\mathrm{rad}\,\mu\mathrm{m}^{-1}$,
$B_1\simeq208.5$~mT, and
$B_2\simeq212.7$~mT. The small difference
$B_2-B_1\simeq4.2$~mT reflects the weak ellipticity and suppresses the parallel-pumping coupling,
\begin{equation}
V_k=
\frac{\gamma}{4}
\frac{B_2-B_1}{\sqrt{B_1B_2}}.
\label{eq:app_pp_V}
\end{equation}
The instability threshold is
\begin{equation}
b_{\parallel,\rm th}
=
\frac{\sqrt{\Gamma_k^2+\Delta_k^2}}{|V_k|},
\label{eq:app_pp_threshold}
\end{equation}
where $b_{\parallel}$ is the microwave-field component parallel to the equilibrium magnetization and
$\Delta_k=\omega_k-\omega_{\rm p}/2$ is the detuning. At exact resonance,
\begin{equation}
b_{\parallel,\rm th}
=
\frac{\Gamma_k}{|V_k|}
\simeq4.2~\mathrm{mT}
\label{eq:app_pp_threshold_res}
\end{equation}
for $\alpha=10^{-4}$.

The threshold given by Eq.~\eqref{eq:app_pp_threshold_res} represents a lower-bound estimate, as only intrinsic Gilbert damping is included. The finite extent of the pumping region, spin-wave propagation beyond this region, inhomogeneous broadening, and incomplete spatial overlap with the pumping field are expected to raise the experimental threshold \cite{Braecher17,Kostylev95}. Below threshold, microwave pumping cannot overcome the relaxation of the $6$~GHz modes, precluding parametric instability. Above threshold, the mode amplitudes are governed by nonlinear saturation and cannot be determined from the threshold condition alone.

In the present geometry, $b_{\parallel}$ corresponds to the out-of-plane microwave-field component. Its amplitude $b_{\rm OP}(x,y)$ was calculated from the time-dependent magnetic field obtained in the TDGL simulations for free vortices at 5 nm below the SC antenna [Fig.~\ref{fig:fig2}(c)]. Using the cosine and sine Fourier coefficients $B_{C,1,z}$ and $B_{S,1,z}$ at the driving frequency of $12$ GHz, the amplitude is given by $b_{\rm OP}(x,y)=\sqrt{B_{C,1,z}^2(x,y)+B_{S,1,z}^2(x,y)}$. The maximum amplitude was found to be $b_{\rm OP,max}\simeq2.05$ mT, below the estimated threshold $b_{\parallel,\rm th}\simeq4.2$ mT. Therefore, the Ga:YIG layer remains below the conventional parallel-pumping instability, and the $12\rightarrow6+6$~GHz process can be excluded as a relevant excitation channel.

\section*{Appendix B: Anharmonicity of vortex motion}
\label{app:vortex_anharmonicity}

To quantify the anharmonicity of the vortex motion shown in Fig.~\ref{fig:fig2}(a), we analyzed the time-dependent mean transverse coordinate $X_v(t)=N_v^{-1}\sum_{j=1}^{N_v}X_{v,j}(t)$ of the $N_v=14$ vortices in the Pb strip. The vortex centers were identified from local maxima of the out-of-plane magnetic-induction component $B_z$, evaluated $5$~nm below the SC strip in the TDGL solutions. Each vortex was tracked from frame to frame, and its position was refined below the grid spacing by fitting a quadratic surface to $B_z$ within a $\pm20$~nm window along $x$ and $\pm1$ grid row along $y$.

The trajectory contains 2502 snapshots over $250$~ps, sampled at slightly nonuniform time intervals of $0.07$--$0.15$~ps (median $0.10$~ps). To suppress the initial transient, we excluded the first driving period, $T=1/f=83.33$~ps, retaining 1666 points over the interval $83.4\leq t\leq249.9$~ps (two periods). Since the sampling times are not equally spaced, we determined the harmonic coefficients by a simultaneous linear least-squares fit, without resampling or a secular drift term, at the fixed fundamental frequency $f=12$~GHz:
\begin{equation}
X_v(t)=X_0+\sum_{n=1}^{10}\left[X_{C,n}\cos(n\omega t)+X_{S,n}\sin(n\omega t)\right],
\label{eq:appB_fit}
\end{equation}
where $\omega=2\pi f$, $X_0$ is the mean position, and $X_{C,n}$ and $X_{S,n}$ are the cosine and sine coefficients, respectively. The peak displacement amplitude $X_n$ and its phase $\phi_n$ are defined by
$X_n e^{i\phi_n}=X_{C,n}-iX_{S,n}$, $X_n=\sqrt{X_{C,n}^2+X_{S,n}^2}$.

The fit gives $X_1=60.1$~nm and $X_3=0.58\pm0.04$~nm ($X_3/X_1\approx1\%$), where the uncertainty reflects the sensitivity to the localization procedure and the vortex-to-vortex variation. The third harmonic is reproducible: separate fits to the second and third driving periods yield $X_3=0.578$ and $0.575$~nm. In contrast, no second harmonic is resolved; its fitted amplitude is below the systematic uncertainty, giving the upper bound $X_2/X_1\lesssim10^{-3}$.
The vortex-center motion is therefore accurately approximated by $X_v(t)\simeq X_0+X_1\cos(\omega t+\phi_1)$. The transverse amplitude is smaller than the $250$~nm half-width of the strip, consistent with confined oscillations. The characteristic maximum vortex velocity estimated from the fundamental harmonic is $\omega X_1\simeq4.5\times10^3$~m/s, approximately one order of magnitude lower than the maximal vortex velocities reported for Pb constrictions in Ref.~\cite{Emb17nac}.

This result helps identify the origin of higher-harmonic spin-wave excitation. In Ga:YIG, the spin-wave amplitude at $2f$ reaches approximately $16\%$ of the fundamental [Fig.~\ref{fig:fig2}(f)], although the second-harmonic vortex-center displacement is at most $\sim0.1\%$ of the fundamental. Thus, the pronounced $2f$ spin-wave signal cannot be explained by a comparably strong second harmonic in the vortex displacement itself. 
% Instead, it supports the spatial-to-temporal transduction mechanism discussed in Sec.~\ref{sec:disc}, in which the nearly sinusoidal vortex motion and the accompanying redistribution of SC currents generate a nonsinusoidal local magnetic field. 
% Instead, it supports the spatial-to-temporal transduction mechanism discussed in Sec.~\ref{sec:disc}, in which the nearly sinusoidal vortex motion drives a nonsinusoidal redistribution of the SC currents and hence a nonsinusoidal antenna field.
Instead, it supports the spatial-to-temporal transduction mechanism discussed in Sec.~\ref{sec:disc}, in which the nearly sinusoidal vortex motion, together with the accompanying redistribution of SC currents, produces a nonsinusoidal antenna field.
The small but resolved $3f$ component of the vortex displacement ($\approx1\%$ of the fundamental) may contribute to the third-harmonic field; its weight relative to the current-redistribution contribution cannot be determined from the trajectory alone.
% We note that $B_z$ maxima are a proxy for the vortex cores; a localization bias independent of the fitting window cannot be excluded from the field data alone. Its quantitative importance cannot be determined from the trajectory alone, because the near-field conversion and the spin-wave susceptibility depend on frequency and position.

\section*{Appendix C: Spectral content of the antenna field and $k_y$ channels}
\label{app:antenna_spectrum}

\begin{figure}[t]
  \centering
  \includegraphics[width=\columnwidth]{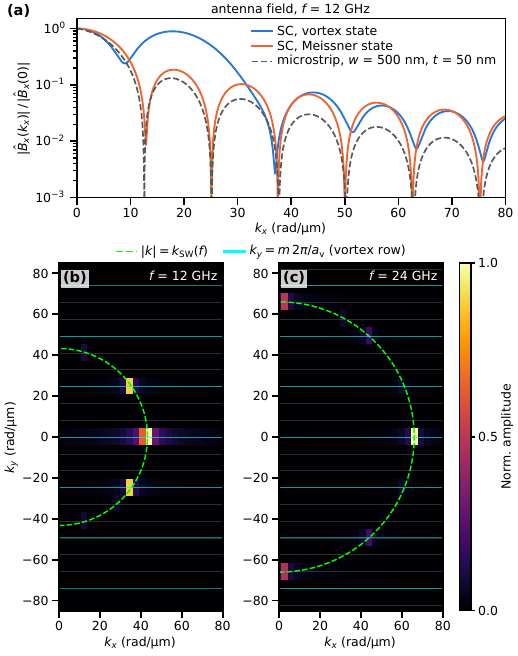}
  \caption{
  (a) Wave-vector spectrum of the $y$-averaged in-plane driving field, $|\hat B_x(k_x)|$, at the excitation frequency $f = 12$~GHz, evaluated 5~nm below the SC strip and normalized to its value at $k_x \to 0$.
  Blue: Pb antenna with freely oscillating vortices at $B_0 = 30$~mT, as used in Figs.~\ref{fig:fig2} and \ref{fig:fig4}; orange: Pb antenna in the Meissner state at $B_0 = 5$~mT, as used in Figs.~\ref{fig:fig3}(f)--\ref{fig:fig3}(j); dashed gray: analytic spectrum of a microstrip [Eq.~\eqref{eq:microstrip_spectrum}] with $w = 500$~nm and $t = 50$~nm.
  (b),(c) Spin-wave amplitude $|\tilde m_x(k_x,k_y)|$ in the Ga:YIG conduit of Fig.~\ref{fig:fig2} at (b) $f$ and (c) $2f$, obtained by a Fourier transform in time and in both in-plane directions over the region $0.5 \le x \le 2.49~\mu$m and normalized to the maximum of each map.
  The dashed green line is the isofrequency contour $|\mathbf k| = k_\mathrm{SW}(f)$ of the forward-volume spin waves.
  Cyan lines mark $k_y = m\,2\pi/a_\mathrm{v}$, the reciprocal-lattice vectors of the vortex row, where $a_\mathrm{v} = 255$~nm is the vortex spacing in the central part of the SC strip; thin gray lines mark $k_y = j\,2\pi/L_y$ set by the periodic simulation cell of width $L_y = 3a_\mathrm{v} = 765$~nm; $m$ and $j$ are integers.}
  \label{fig:SM_antenna_spectrum}
\end{figure}

In the linear regime, the spin-wave emission of an antenna is governed by the Fourier transform of its field, $\hat B_x(k)$, weighted by the gyromagnetic response of the film, which is resonant on the spin-wave dispersion \cite{VlaminckBailleul2010}.
At a given frequency, the response of the film is the same for all antennas, so for equal total current their relative efficiency in exciting spin waves is set by $|\hat B_x(k_\mathrm{SW})|$.
Figure~\ref{fig:SM_antenna_spectrum}(a) compares this spectrum at $f = 12$~GHz for the Pb antenna in the vortex and Meissner states with that of a conventional microstrip of the same width $w = 500$~nm and thickness $t = d_{\rm sc} = 50$~nm.
For a strip of width $w$ and thickness $t$ carrying a uniform current density, the field follows from superposing thin current sheets: each Fourier component of the field of a sheet decays as $e^{-|k||z|}$ with distance $|z|$ from the sheet \cite{VlaminckBailleul2010}, and the Fourier transform of the uniform current across the width is $\propto \sin(kw/2)/(kw/2)$.
Integrating over the strip thickness gives, at a distance $s$ below its lower surface,
\begin{equation}
  \frac{|\hat B_x(k)|}{|\hat B_x(0)|}
  = \left|\frac{\sin(k w/2)}{k w/2}\right| e^{-|k| s}\,
    \frac{1-e^{-|k| t}}{|k| t},
  \label{eq:microstrip_spectrum}
\end{equation}
where the last factor tends to unity for $t \to 0$.

In the Meissner state, the spectrum of the SC antenna follows the sinc-like envelope of Eq.~\eqref{eq:microstrip_spectrum}, with zeros near $k_x = 2\pi n/w$. 
In the vortex state, the $f$ component of $B_x$ above the vortex row oscillates in antiphase with that near the strip edges.
This alternating-sign profile fills the first minimum and strongly enhances the spectral weight at $k_x \approx 15$--$25$~rad/$\mu$m ($\lambda \approx 250$--$420$~nm).
At the wave numbers relevant for the excited spin waves, $k_\mathrm{SW}(f) = 43$ and $k_\mathrm{SW}(2f) = 66$~rad/$\mu$m, both SC spectra lie above the microstrip curve but are nearly equal to each other.
We therefore do not attribute an intrinsic short-wavelength advantage to the vortex state at these $k$.

Because the vortices form a row with period $a_\mathrm{v}$ along $y$, the antenna field also contains components with $k_y = m\,2\pi/a_\mathrm{v}$, where $m$ is an integer.
Each component with $|m|\,2\pi/a_\mathrm{v} < k_\mathrm{SW}(f)$ excites spin waves on the isofrequency contour $|\mathbf k| = k_\mathrm{SW}(f)$ at $k_x = [k_\mathrm{SW}^2 - (m\,2\pi/a_\mathrm{v})^2]^{1/2}$; for $m \neq 0$, these are beams propagating obliquely to the antenna.
The vortex row thus acts as a periodic source, analogous to a diffraction grating: the $m \neq 0$ orders propagate obliquely and, together with the $m=0$ order, form a Talbot-like self-imaging interference pattern \cite{Golebiewski2020,Golebiewski2022}, which gives rise to the $y$-modulation of $m_x$ in Fig.~\ref{fig:fig2}(d).
Figures~\ref{fig:SM_antenna_spectrum}(b) and \ref{fig:SM_antenna_spectrum}(c) show the corresponding $(k_x,k_y)$ maps of the spin-wave amplitude at $f$ and $2f$, respectively.
The strongest signal is located at $k_y = 0$, followed at $f$ by the vortex-row channels $m = \pm1$.
Weaker spots at $k_y = j\,2\pi/L_y$ with $j$ not divisible by 3 (gray lines) result from small differences between the three vortices in the periodic simulation cell, which give the antenna field weak components in these channels.
They are noticeable mainly close to grazing propagation, where the small group velocity along $x$ enhances the spin-wave response.
The $k_y = 0$ channel alone determines the amplitude of the plane wave discussed in the main text.

\section*{Appendix D: SIMULATION DETAILS}
\label{app:Simulations}

\textbf{TDGL simulations.} In the 2D geometry, the generalized time-dependent Ginzburg--Landau (TDGL) equation \cite{Kramer1978,Bishop2023} reads
\begin{equation}
\begin{split}
\frac{u}{\sqrt{1+\Gamma^2|\psi|^2}}
\left(
\frac{\partial}{\partial t}
+i\varphi
+\frac{\Gamma^2}{2}\frac{\partial|\psi|^2}{\partial t}
\right)\psi
={}&(\nabla-i\mathbf{A})^2\psi
\\
&+(1-|\psi|^2)\psi ,
\end{split}
\end{equation}
where $\psi(\mathbf r,t)$ is the complex superconducting order parameter, $u=\pi^4/[14\zeta(3)]\simeq5.79$, $\Gamma=2\tau_\mathrm{E}\Delta_0/\hbar$, $\tau_\mathrm{E}$ is the inelastic electron--phonon scattering time, $\Delta_0$ the zero-field superconducting gap, $\varphi$ the scalar potential, and $\mathbf A$ the vector potential.

The total current density is $\mathbf j=\mathbf j_\mathrm{s}+\mathbf j_\mathrm{n}$, with $\mathbf j_\mathrm{s}=\mathrm{Im}[\psi^\ast(\nabla-i\mathbf A)\psi]$ and $\mathbf j_\mathrm{n}=-\nabla\varphi-\partial\mathbf A/\partial t$, with the current continuity $\nabla\cdot\mathbf j=0$. At SC--vacuum boundaries, $\mathbf n\cdot(\nabla-i\mathbf A)\psi=0$ and $\mathbf n\cdot\nabla\varphi=0$, where $\mathbf n$ is the outward surface normal. At current-injecting SC--normal-metal interfaces, $\psi=0$ and $\mathbf n\cdot\nabla\varphi=j_{\rm ext}$, with $j_{\rm ext}$ the imposed dimensionless current density.

All TDGL quantities are dimensionless. Length and time are scaled by $\xi$ and $\tau_0=\mu_0\sigma_\mathrm{n}\lambda^2$, respectively, where $\sigma_\mathrm{n}$ is the normal-state conductivity and $\xi$ and $\lambda$ are evaluated at the simulation temperature. Defining $B_{\mathrm{c}2}=\Phi_0/(2\pi\xi^2)$ and $j_\mathrm{TDGL}=4\xi B_{\mathrm{c}2}/(\mu_0\lambda^2)$, the vector potential, current density, and scalar potential are scaled by $\xi B_{\mathrm{c}2}$, $j_\mathrm{TDGL}$, and $\xi j_\mathrm{TDGL}/\sigma_\mathrm{n}$, respectively. The order parameter is normalized to its equilibrium magnitude, so that $|\psi|^2$ is the normalized Cooper-pair density.

%To account for the stray field generated by the magnetic layer, we introduce a prescribed
%space- and time-dependent out-of-plane field profile
%$B_z^{\mathrm{FM}}(\mathbf{r},t)$ obtained from micromagnetic calculations.
%The resulting field is
%\begin{equation}
%    B_z(\mathbf{r},t)
%    =
%    B_z^{\mathrm{ext}}
%    +
%    B_z^{\mathrm{FM}}(\mathbf{r},t),
%    \label{eq:tdgl_Bz_total}
%\end{equation}
%where $B_z^{\mathrm{ext}}$ is the out-of-plane component of the external magnetic field applied to the superconductor. The field profile in Eq.~\ref{eq:tdgl_Bz_total} is incorporated into the TDGL equations through the vector potential $\mathbf{A}(\mathbf{r},t)$ by assigning the corresponding local symmetric-gauge values to the TDGL mesh links.
%The magnetic profile is treated as an externally prescribed
%perturbation and is not updated self-consistently by the superconducting currents.

TDGL simulations were carried out for a $4000\times 500$\,nm (length$\times$width) region of a $50$\,nm-thick film subjected to an applied magnetic field $B_0$ of $30$\,mT and $5$\,mT. The material parameters used are summarized in Table~\ref{tab:tdgl_material_p}. The maximum mesh size was set to $\xi/2$. Simulations were performed using the Python package \texttt{pyTDGL}\,\cite{Bishop2023}, with additional custom extensions introducing a space- and time-dependent vector potential.

\textbf{Micromagnetic simulations}. The magnetization dynamics is described by the Landau--Lifshitz--Gilbert equation,
\begin{equation}
\frac{\partial \mathbf{M}}{\partial t}
=
-|\gamma|\,\mathbf{M}\times\mathbf{B}_{\mathrm{eff}}
+
\frac{\alpha}{M_\mathrm{s}}
\mathbf{M}\times
\frac{\partial \mathbf{M}}{\partial t},
\label{eq:LLG}
\end{equation}
where $\mathbf{M}$ is the magnetization, $M_\mathrm{s}$ the saturation magnetization, $\gamma$ the gyromagnetic ratio, and $\alpha$ the Gilbert damping parameter. The effective field $\mathbf{B}_{\mathrm{eff}}$ includes the external, exchange, uniaxial-anisotropy, and dipolar contributions, with the latter evaluated in the magnetostatic approximation \cite{Stancil2009}. The exchange and anisotropy fields are $2A_{\rm ex}M_\mathrm{s}^{-2}\Delta\mathbf{M}$ and $2K_\mathrm{u}M_\mathrm{s}^{-2}(\mathbf{M}\cdot\mathbf z)\mathbf z$, respectively, where $A_{\rm ex}$ is the exchange stiffness, $K_\mathrm{u}$ the perpendicular anisotropy constant, and $\mathbf z$ the out-of-plane unit vector.

Micromagnetic simulations were performed with MuMax3 \cite{Van14aip}. The Ga:YIG conduit was $10.24~\mu\mathrm{m}\times0.765~\mu\mathrm{m}\times20~\mathrm{nm}$ and discretized into $5\times5\times20~\mathrm{nm}^3$ cells. Periodic boundary conditions were applied along $y$ with 128 repetitions. The computational cell along $y$ spans three vortices, while the periodic replicas approximate an extended antenna. A static bias field $B_0$ was applied along $z$, and the equilibrium state was obtained by relaxation. To suppress reflections from the $\pm x$ boundaries, $2~\mu\mathrm{m}$-wide absorbing regions were introduced, with $\alpha$ increasing quadratically to $\alpha_{\max}=0.3$.

The time-dependent magnetic field of the SC antenna was obtained from the TDGL simulations and fitted at each spatial point by a constant term and the first four harmonics of the drive frequency $f=12$~GHz. The reconstructed field used in the micromagnetic simulations was
\begin{equation}
\begin{aligned}
    \mathbf{B}_{\mathrm{app}}(\mathbf r,t)
    ={}&\mathbf{B}_0+\mathbf{B}_{A}(\mathbf r)
    \\
    &+g(t)\sum_{n=1}^{4}
    \left[
    \mathbf{B}_{C,n}(\mathbf r)\cos(n\omega t)
    +
    \mathbf{B}_{S,n}(\mathbf r)\sin(n\omega t)
    \right],
\end{aligned}
\label{eq:hrf}
\end{equation}
where $\omega=2\pi f$, $\mathbf r$ is the position vector, $\mathbf{B}_0$ the uniform bias field, $\mathbf{B}_A(\mathbf r)$ the static antenna-field component, and $\mathbf{B}_{C,n}(\mathbf r)$ and $\mathbf{B}_{S,n}(\mathbf r)$ the cosine and sine spatial coefficients of the $n$th harmonic. The oscillating terms were switched on smoothly using $g(t)=1-e^{-t/\tau}$ with $\tau=30/f=2.5$\,ns. The dynamics was integrated with a fixed time step $\Delta t=1/(100\cdot32f)\approx26$\,fs. After a $50$~ns settling time, the dynamic magnetization component $m_x$ was recorded at 32 frames per drive period over 64 periods. For the $(f,k_x)$ maps, the first eight periods were used, giving a frequency resolution of about $1.5$~GHz; the maximum-amplitude spectra were calculated from all 64 periods, corresponding to about $0.19$~GHz resolution. Prior to spectral analysis, Fourier components with $|k_x|>200~\mathrm{rad}/\mu\mathrm{m}$ were removed. The frequency--wave-vector spectrum was obtained from the Fourier transform of the $y$-averaged space--time map $\langle m_x\rangle_y(t,x)$ within the selected propagation region, where $k_x$ is the wave vector along $x$. The maximum-amplitude spectrum was obtained by Fourier transforming the temporal signal at each spatial point and selecting, for each frequency, the largest Fourier amplitude within the same region.
For the field spectra in Appendix~C, the harmonic field maps of Eq.~\eqref{eq:hrf} were averaged along $y$ and Fourier transformed along $x$.

\bibliographystyle{apsrev4-2}
\bibliography{./main}

% \printbibliography

\end{document}